\pdfoutput=1
\documentclass{JINST}
\usepackage[caption=false]{subfig}
\usepackage{textcomp} % to have the mu in non italics

\newcommand{\se}[1]{\mbox{$#1\,$s}}
\newcommand{\mm}[1]{\mbox{$#1\,$mm}}
\newcommand{\mum}[1]{\mbox{$#1\,$\textmu m}}
\newcommand{\nm}[1]{\mbox{$#1\,$nm}}
\newcommand{\el}[1]{\mbox{$#1\,$e}}
\newcommand{\eV}[1]{\mbox{$#1\,$eV}}
\newcommand{\ke}[1]{\mbox{$#1\,$ke}}
\newcommand{\keV}[1]{\mbox{$#1\,$keV}}
\newcommand{\geVc}[1]{\mbox{$#1\,$GeV/c}}
\newcommand{\uncertainty}[3]{\mbox{$\left(#1\pm#2\right)\,$#3}}

\title{A method for precise charge reconstruction with pixel detectors using binary hit information}
\author{David-Leon Pohl$^a$\thanks{Corresponding author.}~, Jens Janssen$^a$~, Tomasz Hemperek$^a$~, Fabian Hügging$^a$~ and Norbert Wermes$^a$\\
\llap{$^a$}Physikalisches Institut, Universität Bonn,\\
  Nu\ss allee 12, 53115 Bonn, Germany\\
  E-mail: \email{David-Leon.Pohl@uni-bonn.de}}

\abstract{A method is presented to precisely reconstruct charge spectra with pixel detectors using binary hit information of individual pixels. The method is independent of the charge \mbox{information} provided by the readout circuitry and has a resolution mainly limited by the \mbox{electronic} noise. It relies on the ability to change the detection threshold in small steps while counting hits from a particle source.
The errors are addressed and the performance of the method is shown based on measurements with the ATLAS pixel chip FE-I4 bump bonded to a \mum{230} 3D-silicon sensor. Charge spectra from radioactive sources and from electron beams are presented serving as \mbox{examples}. It is demonstrated that a charge resolution ($\sigma$~<~\el{200}) close to the electronic noise of the ATLAS FE-I4 pixel chip can be achieved.
}
\keywords{pixel detectors; charge reconstruction; method; charge resolution; ATLAS FE-I4}

\begin{document}
%
% Section Motivation
%
\section{Motivation}
Pixel detectors are used in particle physics experiments where high track densities require tracking detectors with good spacial resolution combined with fast readout and time stamping capabilities~\cite{bib1}. The trend in many experiments to have successively higher luminosities requires new pixel detectors with smaller pixel sizes and faster processing times to avoid pile-up in pixels. The optimization for high hit rates with reasonable power consumption comes at a cost of charge resolution~\cite{bib3}. This renders the calibration and characterization of the pixel detector challenging. Therefore a charge reconstruction-method was developed that is independent of the charge information provided by the detector. The method depends solely on the digital information whether a pixel is hit and has a resolution mainly limited by electronic noise. The incentive for the development of this method came from the need to carry out charge measurements with the ATLAS pixel chip FE-I4, which only has very limited charge resolution. The method is, however, restricted to the charge reconstruction of individual pixels instead of pixel clusters.
%
% Section Method
%
\section{Method}
\begin{figure}[ht]
	\centering
		\includegraphics[width=1.0\textwidth]{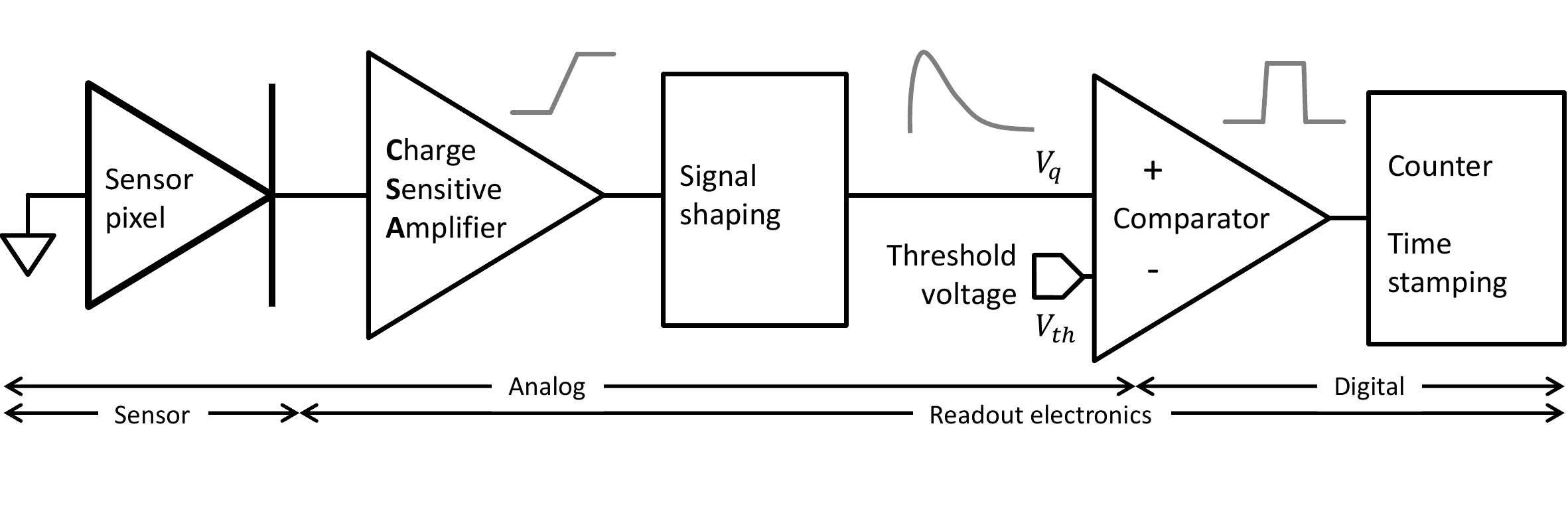}
	\caption{Typical readout chain of one pixel of a counting pixel detector.}
\label{fig:Readout_scheme}
\end{figure}

\begin{figure}[hb]
	\centering
		\includegraphics[width=0.63\textwidth]{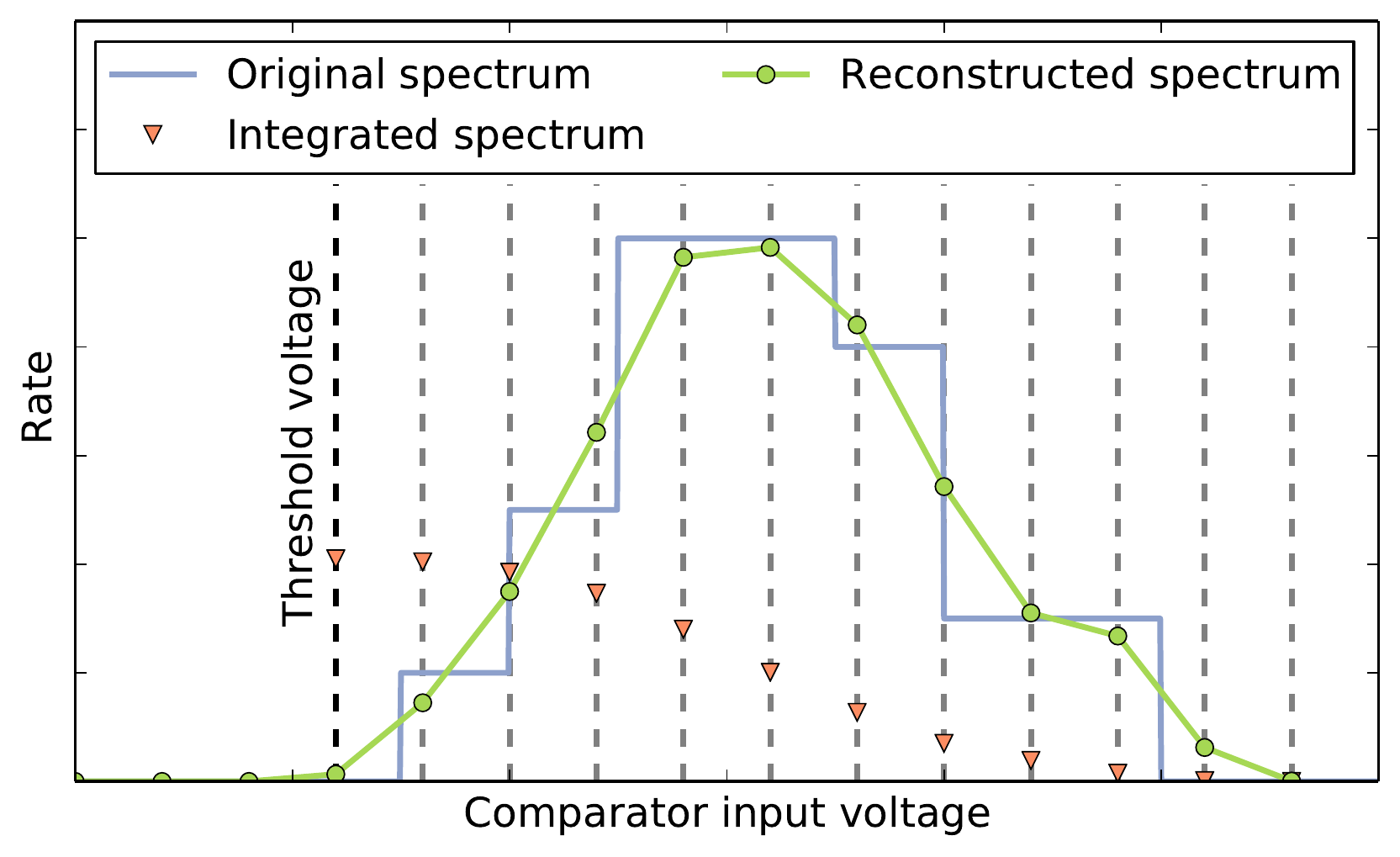}
	\caption{The threshold method illustrated by an arbitrary pulse amplitude spectrum (\textit{blue line}). The different threshold positions (\textit{grey dotted line}), the integrated spectrum (\textit{red triangle dots}) and the reconstructed spectrum (\textit{green circle dots}) are shown. The electronic noise is half of a threshold step.}
\label{fig:method}
\end{figure}

The typical readout chain of one pixel in a pixel detector is shown in Figure~\ref{fig:Readout_scheme}.
The depleted sensor pixel, depicted as a reversed biased diode, is followed by the readout electronics consisting of a charge sensitive amplifier (CSA), a shaping stage, and a comparator. The output of the shaping stage is a voltage pulse whose amplitude corresponds to the charge deposited in the pixel. The amplitude $V_q$ is compared to a configurable threshold voltage $V_{th}$ resulting in a digital signal at the output of the comparator. The digital signal indicates that a pixel is hit and is used by the digital logic for further processing, e.g. time stamping and counting. The threshold voltage~$V_{th}$ is usually finely adjustable on pixel level since it defines the hit detection-threshold and has a big impact on the homogeneity and the signal-to-noise performance of the pixel detector. The basic idea of the method presented in this paper is to use the adjustable threshold voltage $V_{th}$ to change the hit detection threshold while measuring the hit rate of a particle source. This leads to a recording of the integrated pulse amplitude $V_q$ spectrum from which the original charge spectrum can be deduced. This method (henceforth referred to as \textit{threshold method}) is shown in Figure~\ref{fig:method}.
At first, a threshold  voltage above the highest occuring signal amplitude is set. Then the thresold voltage is successively reduced while simultaneously recording the corresponding hit rate (\textit{red triangles}). The integrated spectrum is differentiated leading to the reconstructed spectrum (\textit{green circles}). This represents the original pulse amplitude-spectrum (\textit{blue line}) convoluted with a normal distribution, whose standard deviation is given by the electronic noise.
To deduce the single pixel charge spectrum from the reconstructed pulse amplitude spectrum a calibration is needed, because usually the signal amplitude $V_q$ as a function of the charge and the threshold voltage $V_{th}$ are unknown. The calibration can be done with a charge injection circuit (Section~\ref{sec:setup},~Figure~\ref{fig:threshold_lookup}) or different gamma sources (Section~\ref{sec:results},~Figure~\ref{fig:calibration}).
%Usually neither the threshold voltage $V_{th}$ as a function of the threshold setting nor the signal amplitude $V_q$ as a function of the charge is known. Therefore a calibration describing the detection threshold for different charges is needed (Section~\ref{sec:setup},~Figure~\ref{fig:threshold_lookup}). With this calibration the single pixel charge spectrum can be deduced from the reconstructed pulse amplitude spectrum. The calibration can be done with different gamma sources (Section~\ref{sec:results},~Figure~\ref{fig:calibration}) or a charge injection circuit.

\noindent The requirements for the threshold method to reconstruct a charge spectrum $q \in [q_{min}, q_{max}]$ are:
\begin{eqnarray}
  \min(V_{th}) &<& V(q_{min}) - 2.35\ V_{noise}\label{eq:requirements1}\nonumber \\
  \max(V_{th}) &>& V(q_{max}) + 2.35\ V_{noise}\label{eq:requirements2}\label{eq:requirements}\\
  V(q) && \mathrm{strictly~monotonic}\ \forall\ q \in [q_{min}, q_{max}]\nonumber 
\end{eqnarray}
$V_{noise}$ is the standard deviation of the electronic noise. The requirements state that the threshold voltage range has to be sufficiently large to resolve the charge spectrum. The minimum/maximum threshold voltage has to be one Full-Width-Half-Max (FWHM$\ \approx2.35\sigma$) apart from the minimum/maximum signal amplitude. The strict monotonicity requirement of $V(q)$ demands that the CSA does not saturate.
%
% Section Setup
%
\section{Experimental setup}\label{sec:setup}
\begin{figure}[b!]
	\centering
		\includegraphics[width=0.85\textwidth]{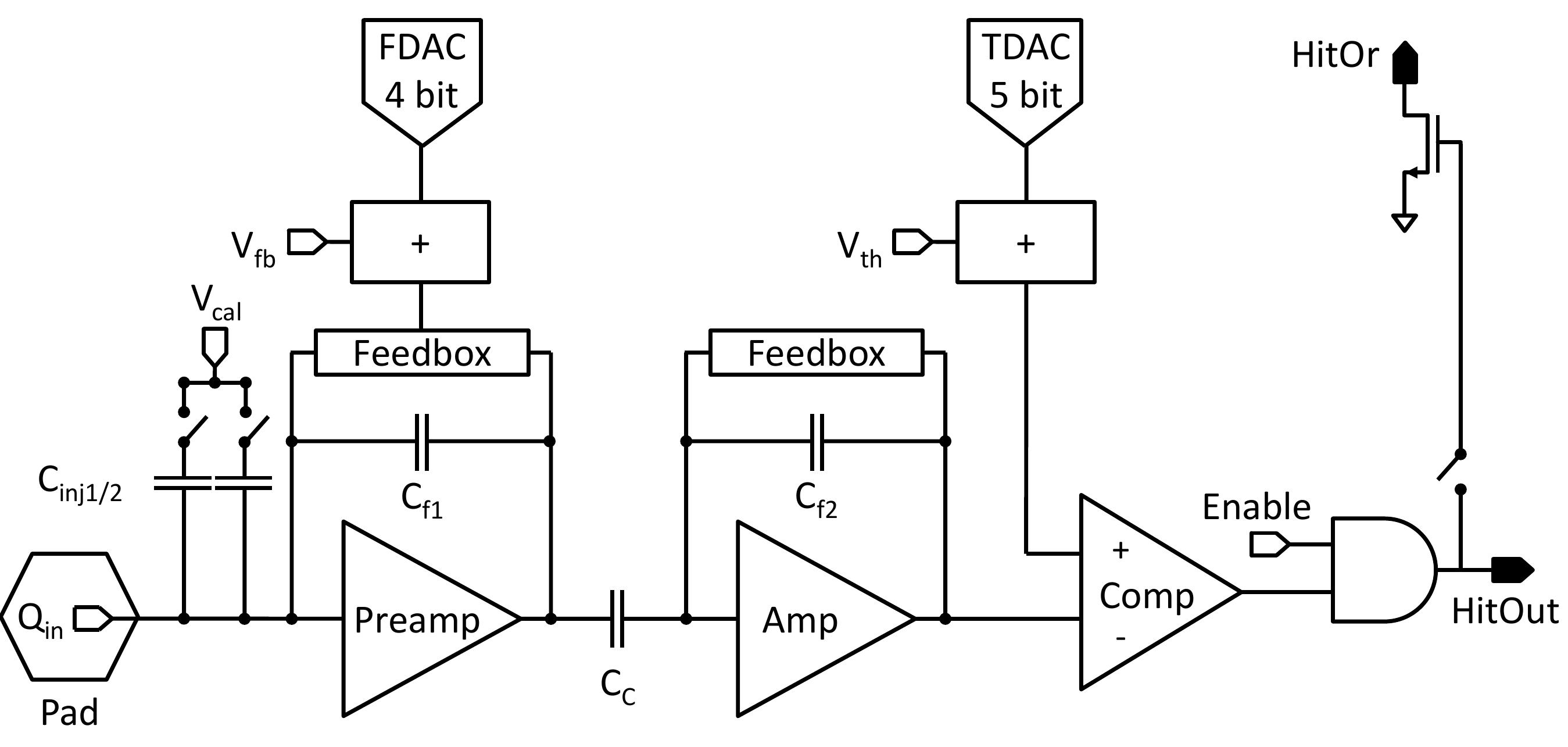}
	\caption{Simplified schematic of the analog pixel cell of the ATLAS FE-I4 consisting of two AC coupled charge sensitive amplifiers (\textit{Preamp/Amp}), a comparator (\textit{Comp}), a charge injection circuit ($V_{cal}$), shaping time and threshold tuning circuits ($V_{fb}, FDAC; V_{th}, TDAC$), an enable switch (\textit{Enable}), and a HitBus signal~(\textit{HitOr}).}
\label{fig:analog_pixel}
\end{figure}

The measurements were carried out with a prototype assembly of a hybrid pixel detector. The assembly was developed within the scope of the ATLAS pixel detector-upgrade~\cite{bib4}. It consists of 26,880 pixels with as size of \mum{250}~x~\mum{50} arranged in an 80~x~336~array~\cite{bib5}. A CNM 3D-silicon sensor was used with a sensitive area of \mm{18.75}~x~\mm{20.45} and \mum{230} thickness.
In contrast to the planar sensor design, the 3D-sensor electrodes are etched into the bulk from both sides as columns of \mum{10} radius~\cite{bib6}. Each pixel has two n-type readout electrodes and six p-type biasing electrodes shared with neighboring cells. The readout is done with the ATLAS pixel readout-chip Front-End-I4 (FE-I4A), an integrated circuit in \nm{130} CMOS technology~\cite{bib7}. The Front-End pixels are connected to the sensor pixels via Sn-Ag bump bonds. Each Front-End pixel has a two staged AC-coupled CSA with adjustable shaping time for the first stage and an adjustable threshold for the discriminator that follows the second stage (Figure~\ref{fig:analog_pixel}).
A feedback and threshold DAC~(\textit{FDAC/TDAC}) present in each pixel allow fine tuning of the individual shaping time and threshold. The charge determination and time stamping is implemented using the time-over-threshold technique with 4-bit resolution. The time-over-threshold is given in counts of an externally supplied clock of $40$\,MHz. The comparator output of each pixel (\textit{HitOut}) can be ORed to form a hit bus~(\textit{HitOr}). The signal of the hit bus can be used to start an automatic read out without an external trigger (FE-I4 self-trigger mode). The FE-I4 also holds an internal charge injection circuit for tuning and testing. It distributes a voltage step ($V_{cal}$) to selectable injection capacitors ($C_{inj1},\ C_{inj2}$) present in each pixel. The absolute charge injected by the circuit can be determined with $\sim$~10\% uncertanty, because the pulse amplitude and injection capacitor values are only roughly known~\cite{bib20}.
\begin{figure}[h]
	\centering
		\includegraphics[width=0.63\textwidth]{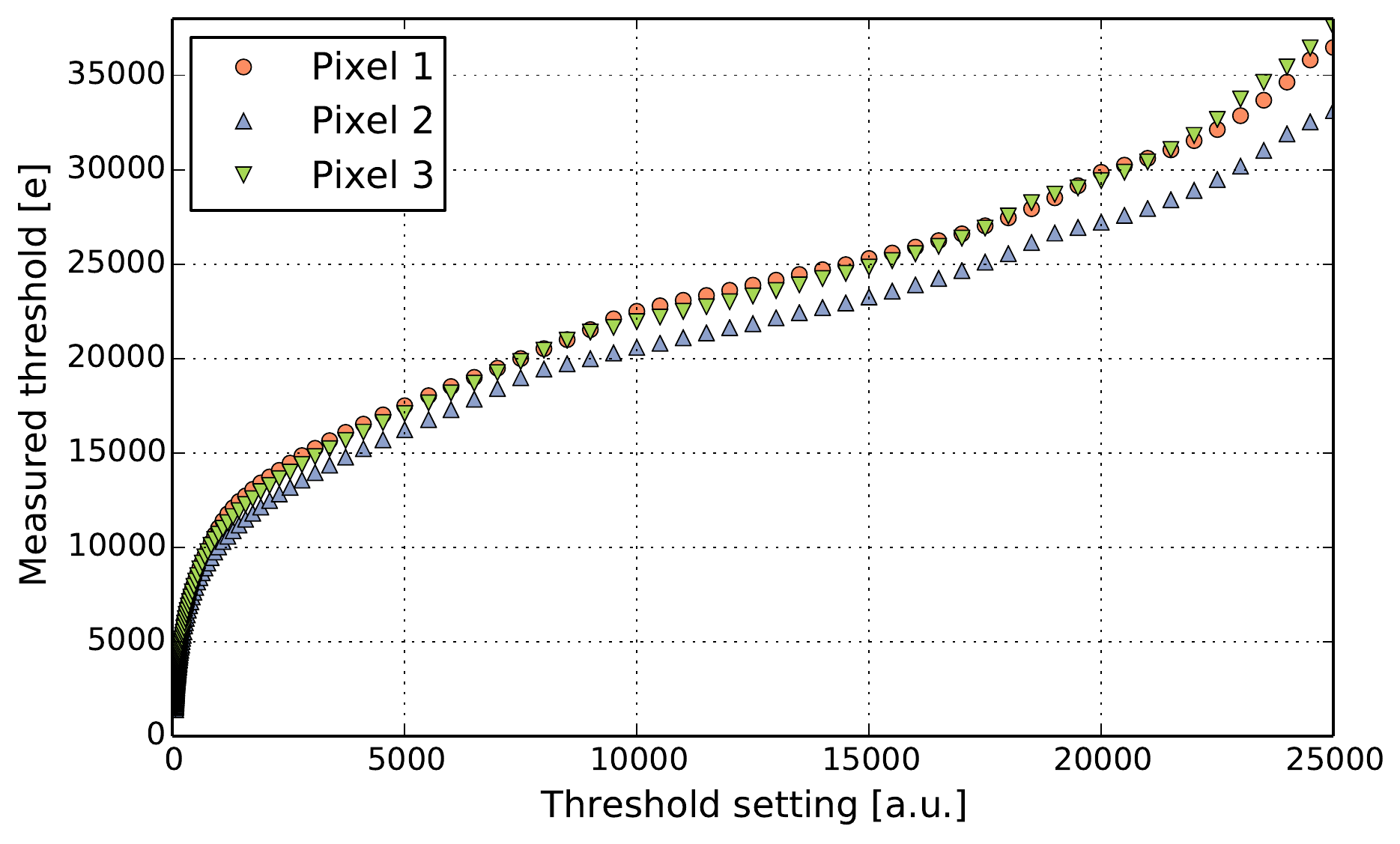}
	\caption{Threshold calibration curves of the FE-I4 pixel readout chip. Three different pixels are depicted. The x-axis shows the threshold DAC setting and the y-axis the measured threshold level in electrons. The error of the measured threshold is with \el{< 10} very small and therefore not shown.}
\label{fig:threshold_lookup}
\end{figure}

\noindent The hybrid pixel assembly was glued onto a printed circuit board (PCB), with wire-bonds connecting the signal and power pads of the Front-End to the PCB. The communication with the Front-End was done with USBpix, an FPGA based versatile readout system for detector tests~\cite{bib8}\cite{bib9}. 
Data aquisition and analysis was performed using pyBAR\footnote{pyBAR - \textbf{B}onn \textbf{A}TLAS \textbf{R}eadout in \textbf{Py}thon}, a readout software written in Python~\cite{pybar}. All measurements were carried out with a fully depleted CNM 3D-sensor with 20~V bias at room temperature. The FE-I4 was tuned to \el{\sim 2000} threshold with \el{\sim 50} pixel-to-pixel dispersion. The electronic noise was \uncertainty{160}{12}{e}. The threshold dispersion is smallest at the position of the threshold tuning and diverges if the global threshold (Figure \ref{fig:analog_pixel}, $V_{th}$ ) is changed. A threshold tuning at high thresholds (\ke{>10}) is not possible, due to range limitations of the pixel threshold DACs~(\textit{TDAC}). However, for the threshold method it is crucial to know the exact threshold for each pixel. Therefore the threshold was measured with the internal charge injection circuit for each pixel and for several threshold settings. A calibration in the form of a lookup table was created~(Figure \ref{fig:threshold_lookup}).
Figure~\ref{fig:threshold_lookup} also shows that the monotone requirement (\ref{eq:requirements}) for charges up to \el{35000} is fulfilled. To retain the threshold calibration validity the voltages for the Front-End were sensed to mitigate any voltage drop on the power cables that significantly affects the threshold position. The radioactive source measurements were carried out in the FE-I4 self-trigger mode as a table top experiment. The electron beam data were recorded at test beams at DESY\footnote{\textbf{D}eutsches \textbf{E}lektronen-\textbf{Sy}nchrotron} in Hamburg and at ELSA\footnote{\textbf{El}ektronen-\textbf{S}trecher-\textbf{A}nlage} in Bonn with \geVc{3.0 - 3.5} beam momentum. The trigger for the detector during beam measurements was provided by a trigger logic unit (TLU) that has been developed within the EUDAQ/AIDA framework~\cite{tlu}\cite{aida}. Two scintillators were placed before and after the assembly and a trigger was only issued if both were hit in coincidence. Neither the electron beam nor the source particles completely illuminated the sensor. Therefore a position in the pixel matrix was chosen, where the threshold dispersion after chip tuning is smallest.
%
% Section Errors
%
\section{Errors}\label{sec:errors}
\begin{figure}[h]
	\centering
	\subfloat[][]{\label{fig:Event_rate}\includegraphics[width=0.5\linewidth]{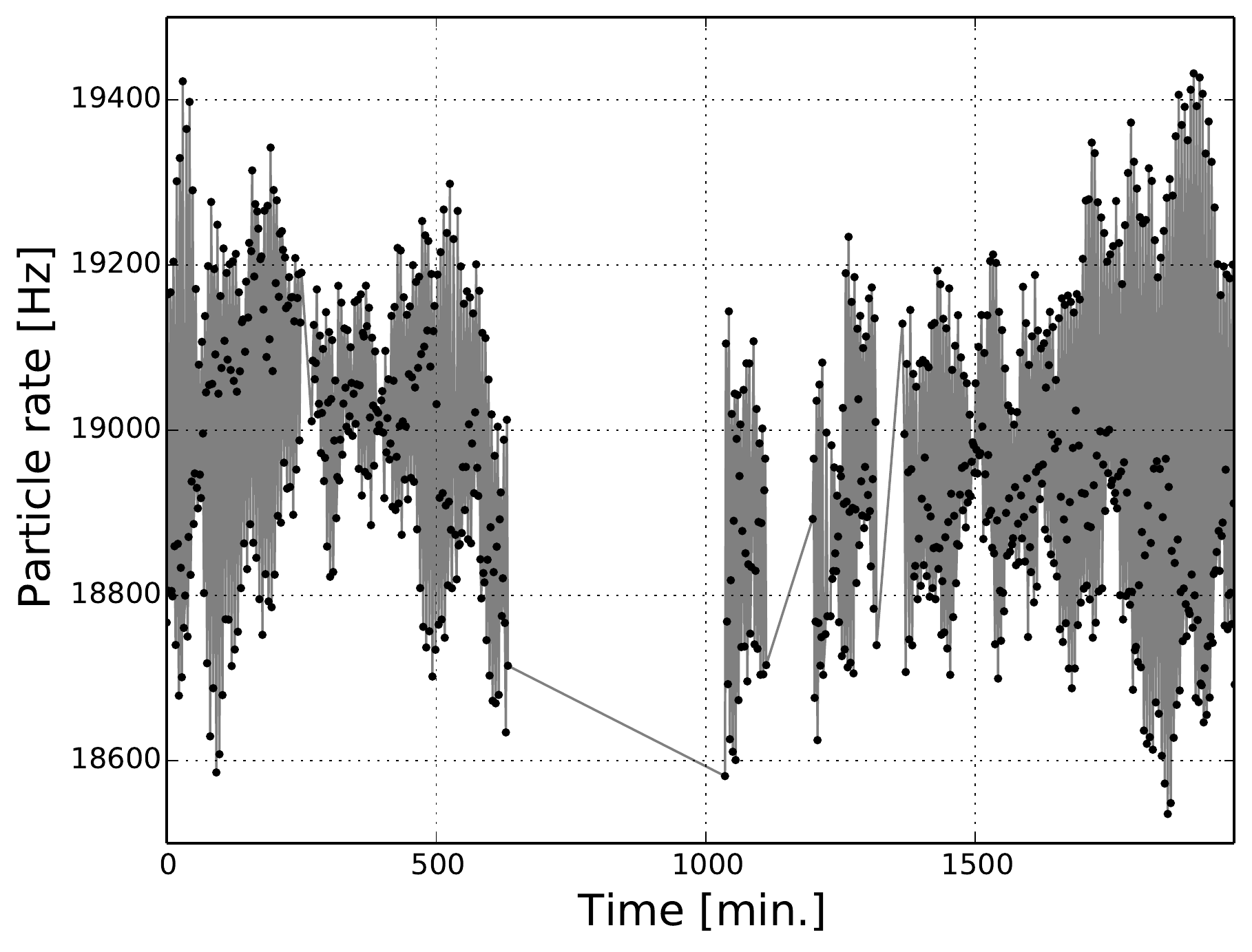}}
	\subfloat[][]{\label{fig:multiplicity}\includegraphics[width=0.49\linewidth]{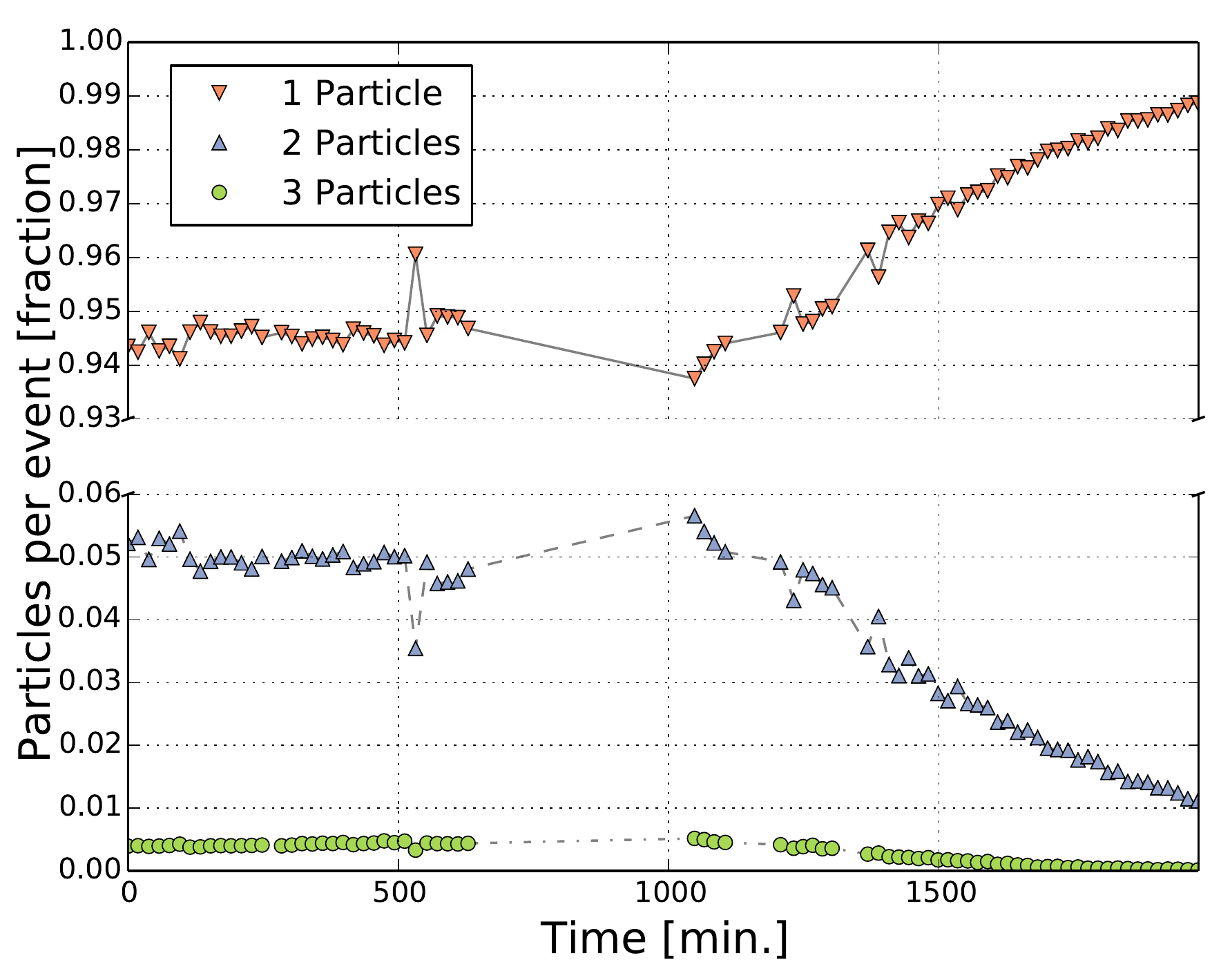}}
	\caption{The measured particle rate (a) and the number of particles per event (b) measured at the DESY testbeam facility over a period of 33~hours. The gaps are periods without beam. One data point in (a) is the average particle frequency for \se{\sim 100}. One point in (b) represents the data averaged over 24~min. The particle rate and the particles per event changed about 5~\% with time.}
	\label{fig:beam_rate}
\end{figure}

The threshold method works under the assumption that the change of the single pixel rate arises solely from the different threshold settings. All other effects influencing the rate have to be measured and substracted out. The prerequisite that the particle rate of the source is constant is only given for a radioactive source where the activity can be assumed constant. In a particle accelerator experiment the beam conditions are often not in control of the experimenter and can change with time. Therefore, a triggered detector readout system or an independent measurement of the delivered particle rate (e.g. with scintillators) is needed to correct the measured rate (Figure~\ref{fig:Event_rate}).
Another effect changing the measured rate is a change of the number of particles per event (Figure~\ref{fig:multiplicity}).
A correction for this can be done by analyzing the number of clusters per event and applying a correction factor for each threshold setting to the data.
Both effects depend on the accelerator and have been determined at the test beam facilities at DESY and ELSA to change the measured rate by about 10\%.

\begin{figure}[t]
	\centering
		\includegraphics[width=0.63\textwidth]{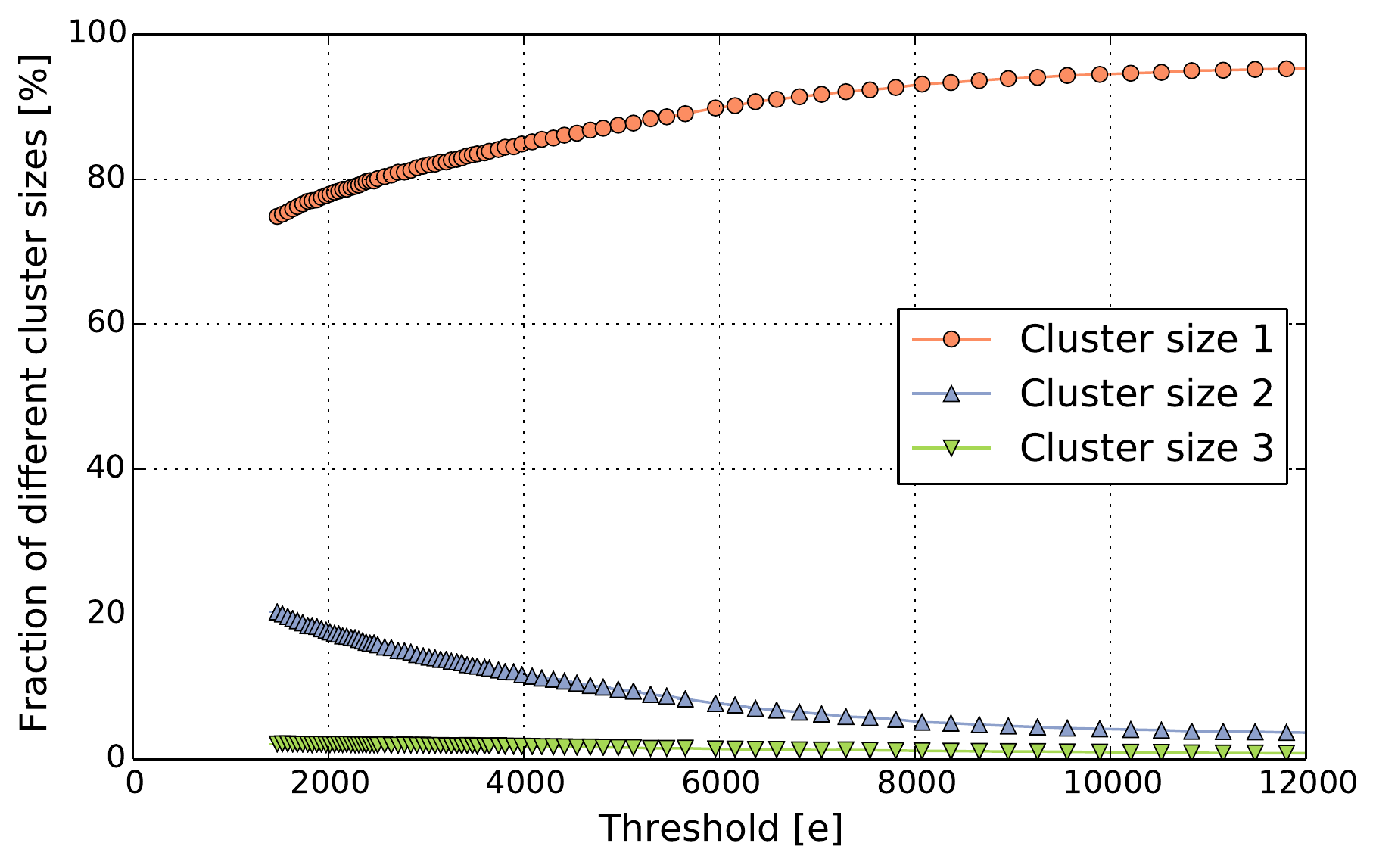}
	\caption{The fraction of 1, 2, and 3-hit clusters as a function of the mean threshold. Data obtained with an ATLAS FE-I4 pixel chip on a CNM 3D-sensor in a \geVc{3.4} electron beam.}
\label{fig:Cluster_sizes}
\end{figure}

\begin{figure}[b]
	\centering
		\includegraphics[width=0.63\textwidth]{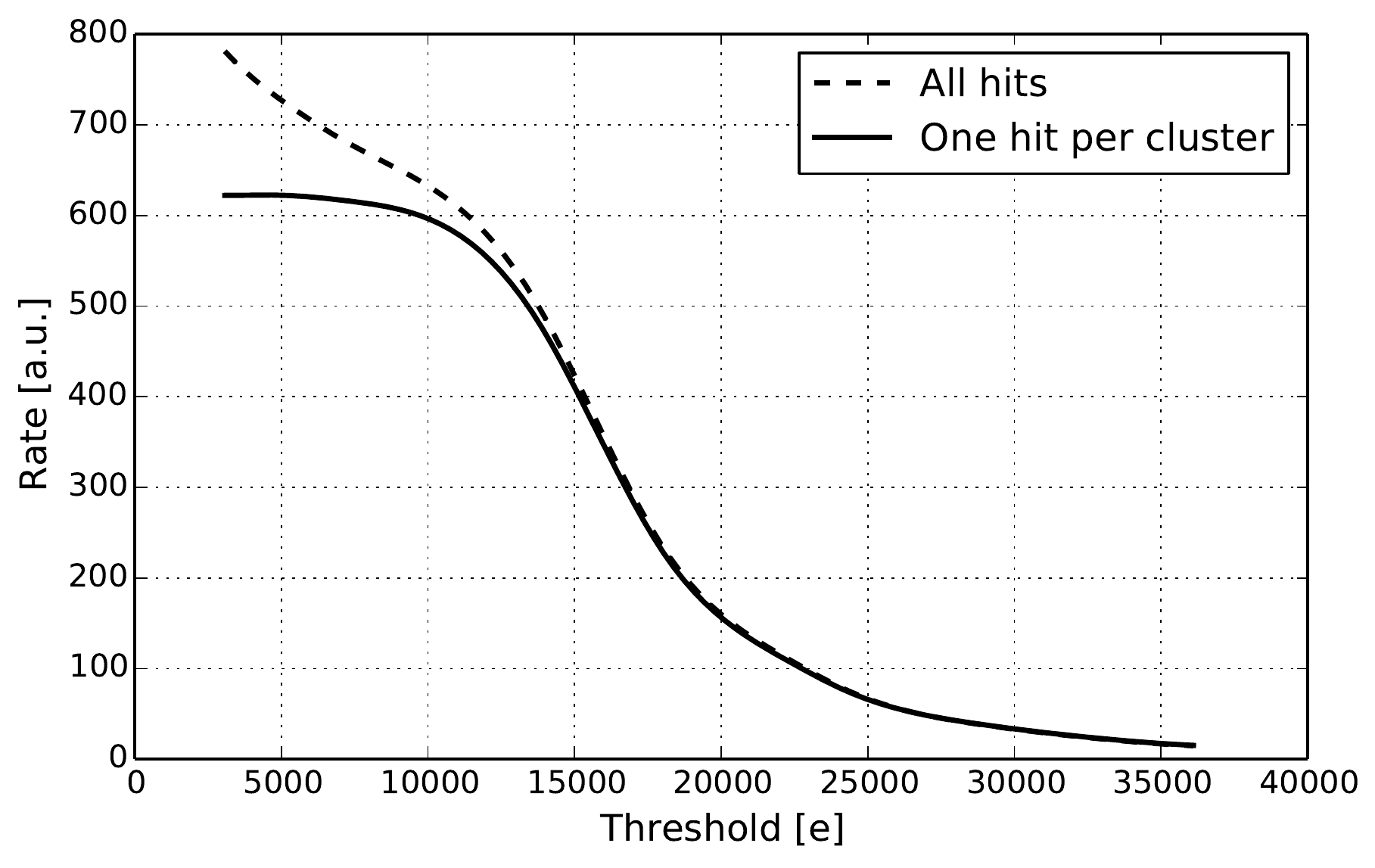}
	\caption{Integrated single pixel charge spectrum measured using the threshold method for all hits taken into account (\textit{dashed line}) and one hit per cluster (\textit{dashed line}). If only one hit per cluster is used the rate does not increase for lower thresholds due to larger cluster sizes. Data obtained with an ATLAS FE-I4 pixel chip on a CNM 3D-sensor in a \geVc{3.2} electron beam.}
\label{fig:Cluster_correction}
\end{figure}

\begin{figure}[h]
	\centering
		\includegraphics[width=0.60\textwidth]{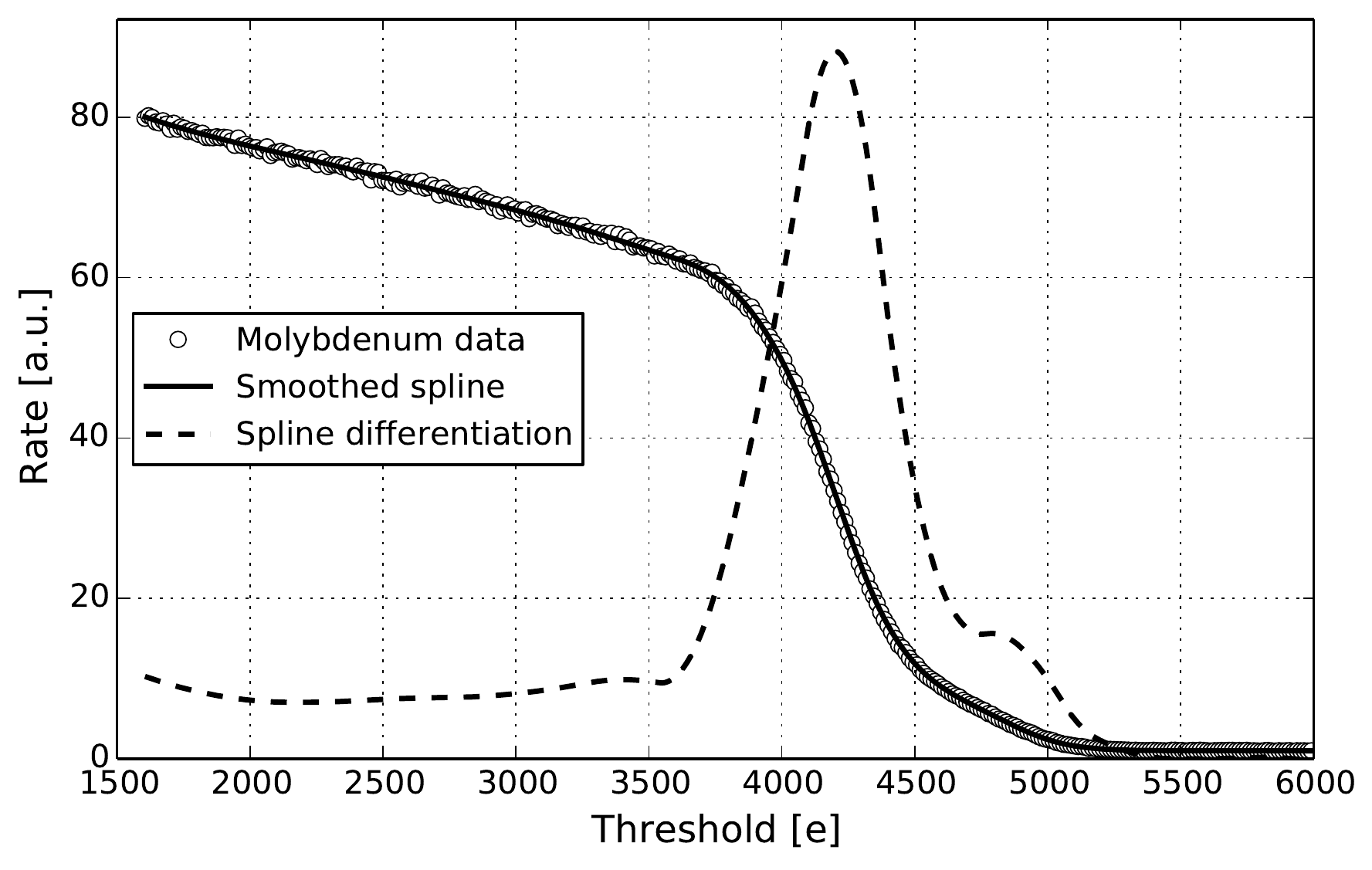}
	\caption{The integrated and differentiated X-ray spectrum of Molybdenum recorded using the threshold method. One \textit{point} corresponds to one bin of a profile histogram combining the data of $\sim 5000$~pixels. The \textit{black line} shows the fit of a spline curve of 3rd order including errors. The spline differentiation is depicted as a \textit{dotted line}.}
\label{fig:example_differentiation}
\end{figure}

\noindent So far only single pixel hits were considered. Often a particle deposits charge in several neighboring pixels due to charge sharing and multiple scattering. With decreasing threshold it is more likely to see not only the seed pixel hit but also additional hits from its neighbors~(Figure~\ref{fig:Cluster_sizes}). 
As a result the measured single pixel hit rate increases artificially with lower thresholds. Considering only one random hit per pixel cluster corrects for this effect, because this leads to particle counting instead of hit counting. The integrated single pixel charge spectrum before and after correction is depicted in Figure~\ref{fig:Cluster_correction}. The expected S-curve like shape arising from the integrated Landau spectrum can be seen if only one hit per cluster is used.
The differentiation of the measured integrated spectrum is mathematically not straight forward. Conventional finite-difference approximations greatly amplify the statistical fluctuations in the measured rate~\cite{bib14}. This can be mitigated in two ways. Illuminated pixels can be combined to increase the statistics and consequently decrease the statistical fluctuations. Further the data can be denoised before differentiation. The data of illuminated pixels have therefore been combined using profile histograms (Figure~\ref{fig:example_differentiation}). The denoising and differentiation was done with smoothed spline curve fits of 3rd order \cite{bib15},~\cite{bib16} provided by the SciPy Python package~\cite{scipy}.
%
% Section Reults
%
\section{Results}\label{sec:results}
\begin{figure}[t]
	\centering
		\includegraphics[width=0.63\textwidth]{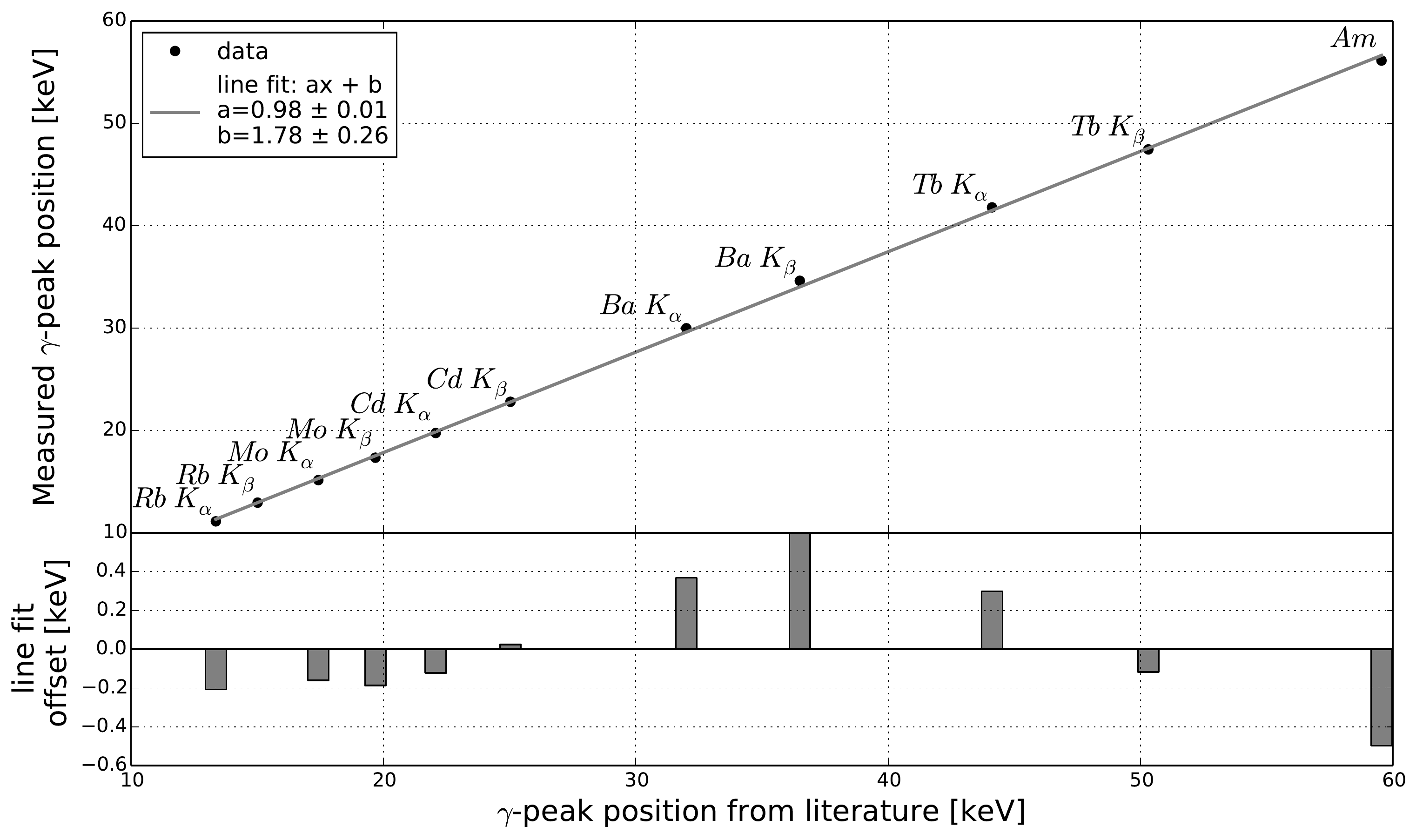}
	\caption{The measured gamma peak positions of the $K_\alpha$ and $K_\beta$ transitions of Rubidium~(\textit{Rb}), Molybdenum~(\textit{Mo}), Cadmium~(\textit{Cd}), Barium~(\textit{Ba}) and Terbium~(\textit{Tb}). The \textit{Am} point marks the most propable gamma transition of the decay product $^{237}Np$ of $^{241}Am$. The literature values where taken from~\cite{radionucleides},~\cite{xray}. Top: The data is fitted with a straigth line (\textit{gray line}). Bottom: The offset of the line fit for each data point is depicted as bars.}
\label{fig:calibration}
\end{figure}

Multiple single pixel charge spectra were recorded using the threshold method to determine its performance and feasibility.  
\begin{figure}[h!]
	\centering
	\subfloat[][]{\label{fig:Rubidium}\includegraphics[width=0.49\linewidth]{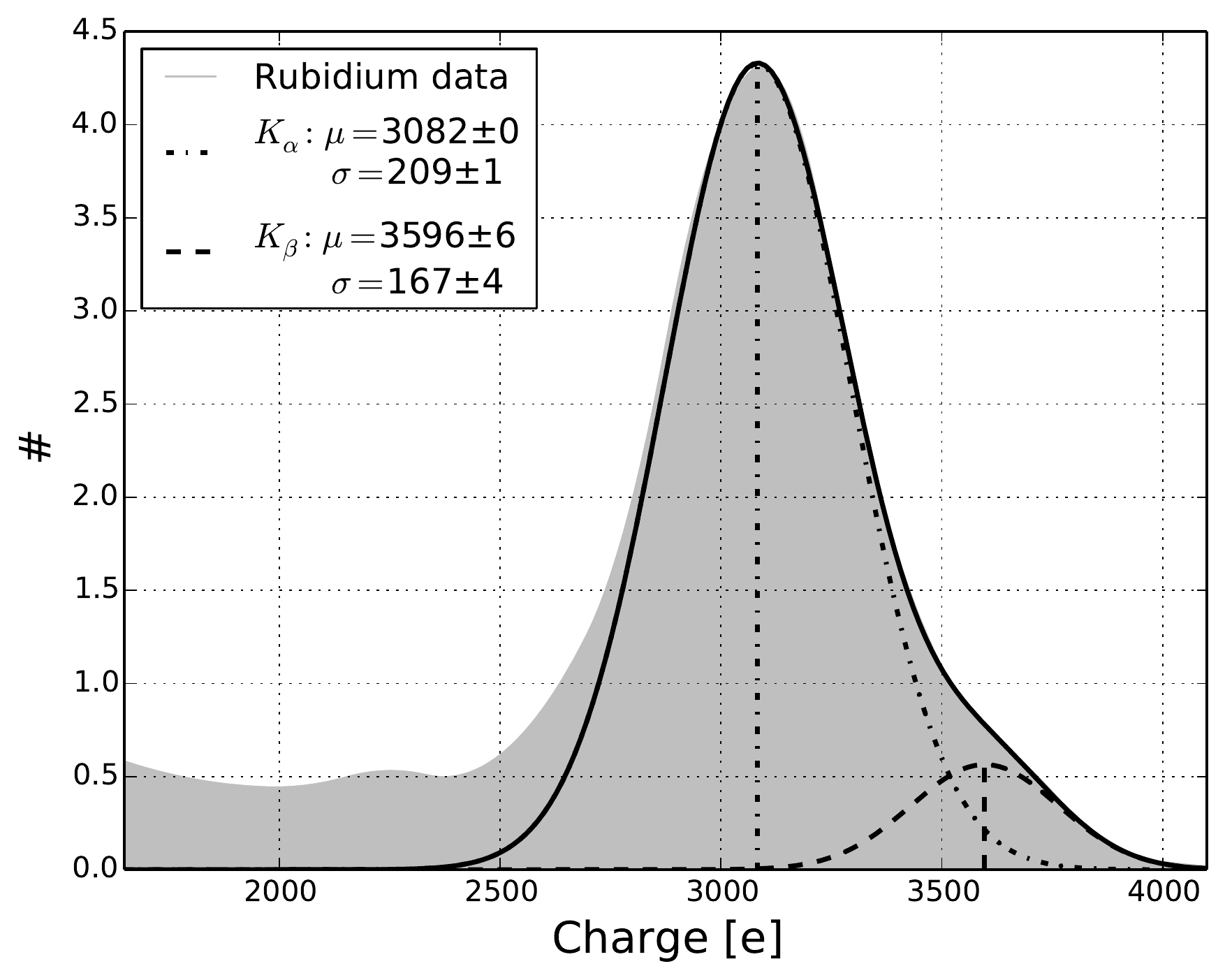}}
	\subfloat[][]{\label{fig:Cadmium}\includegraphics[width=0.49\linewidth]{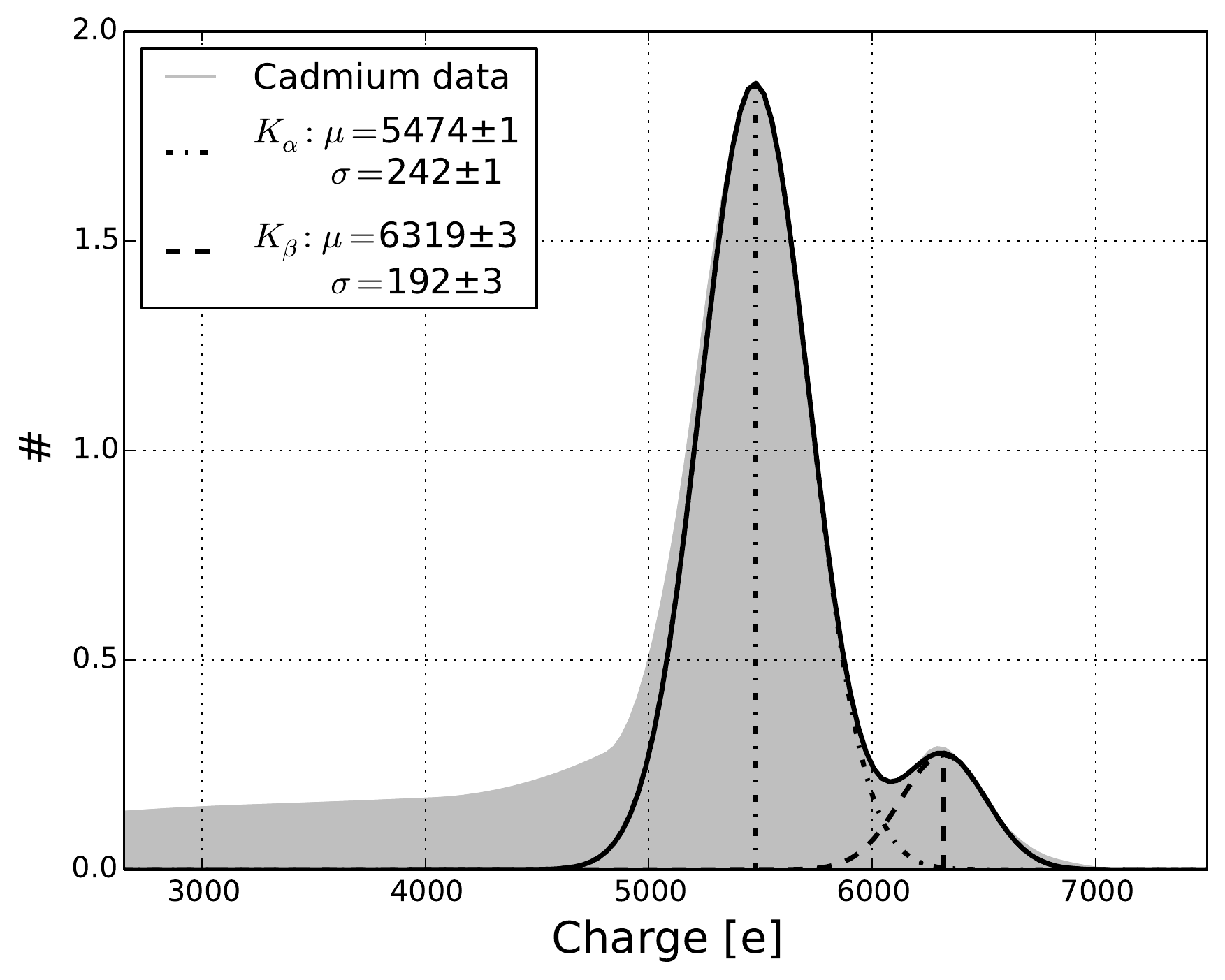}}
	\caption{X-ray spectra of Rubidium~(\textit{a}) and Cadmium~(\textit{b}) measured using the threshold method. The $K_\alpha$ and $K_\beta$ emission lines were fitted with the sum of two normal distribution functions (\textit{black lines}).}
\label{fig:xray_spectra}
\end{figure}
Charge spectra from X-Ray emissions of two elements are depicted in Figure~\ref{fig:xray_spectra}. The x-axis shows the threshold position in electrons. An error for the threshold is expected since its value relies on calibration constants for the charge injection circuit of the device. These constants are known with limited precision only~(Section~\ref{sec:setup}). To calibrate the threshold position several $K_\alpha$ and $K_\beta$ emission lines of different elements were fitted with the normal distribution function and plotted against their literature values (Figure~\ref{fig:calibration}).
The conversion from charge to energy was done assuming that \eV{3.61} are needed to create one electron-hole pair in silicon~\cite{electronholes}.
\begin{figure}[b]
	\centering
		\includegraphics[width=0.63\textwidth]{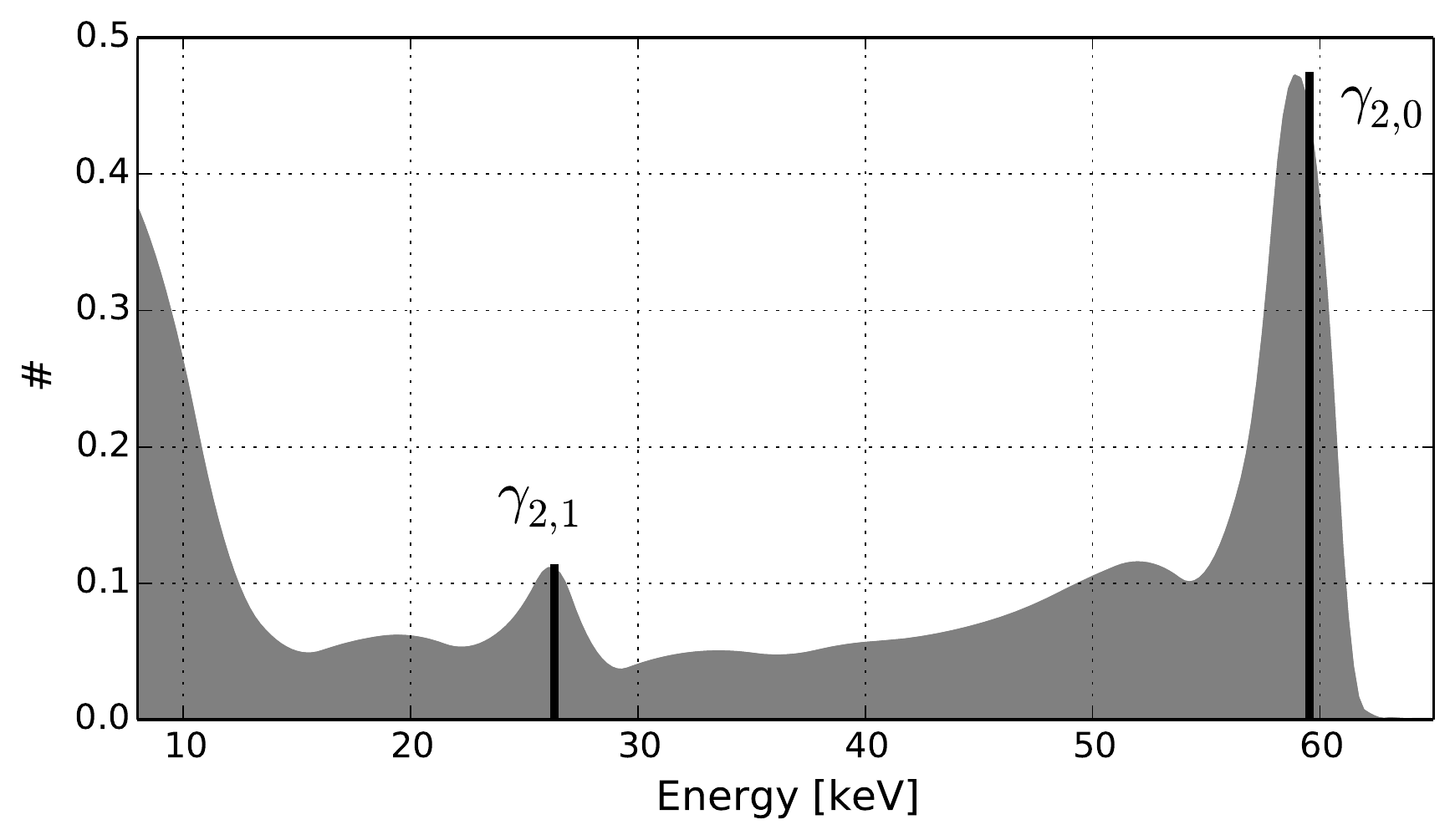}
	\caption{Americium-241 gamma spectrum measured with the threshold method. The most probable gamma transitions within the scan range of the decay product Np-237 are highlighted: $\gamma_{2,0}=59.5~\mathrm{keV}$ and $\gamma_{2,1}=26.3\,\mathrm{keV}$ (\textit{black lines})~\cite{radionucleides}.}
\label{fig:am_spectrum}
\end{figure}
The deviations from the line fit (Figure~\ref{fig:calibration},~bottom) arise from divergences in the pixel-per-pixel threshold lookup-table (Figure~\ref{fig:threshold_lookup}) and the spline smoothing of the measured integrated spectra. The maximum deviation defines the maximum error of the threshold method with the used device and is rather small with \keV{0.6}~$\widehat{=}$~\el{165}.
The minimum standard deviation of the normal distributions of the X-Ray emission lines in Figure~\ref{fig:xray_spectra} is \el{\sim 190}. This demonstrates that the resolution of the threshold method is limited mainly by the electronic noise of the device with \uncertainty{160}{12}{e}.
\begin{figure}[t]
	\centering
		\includegraphics[width=0.63\textwidth]{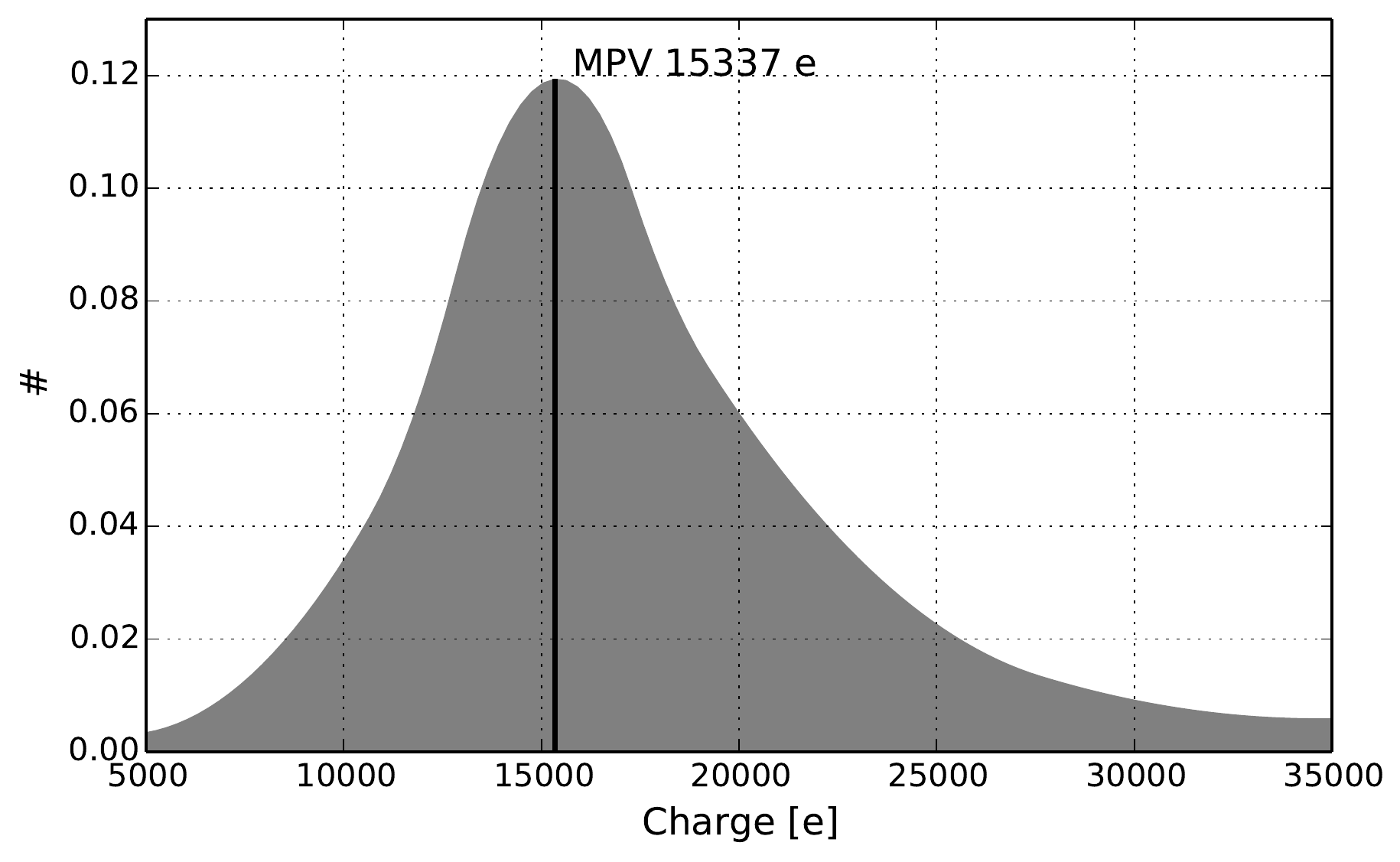}
	\caption{The single pixel charge spectrum of \geVc{3.4} electrons in a \mum{230} 3D silicon sensor measured using the threshold method. The most propable value (MPV) is shown by a \textit{black line}.}
\label{fig:landau}
\end{figure}
Also charge spectra with many features can be resolved with the threshold method as depicted by the Americium spectrum in Figure~\ref{fig:am_spectrum}. The most probable gamma transitions from literature within the scan range are given by black lines~\cite{radionucleides}. They coincide with the data if the threshold calibration~(Figure~\ref{fig:calibration}) is applied. The charge spectrum of a \geVc{3.4} electron beam is depicted in Figure~\ref{fig:landau}. The most probable value (MPV) of the Landau charge-distribution for minimum ionizing particles in \mum{230} silicon is expected to be at \el{16,200}~\cite{eloss}. The determined MPV with \el{\sim 15,350} is slightly lower since the threshold method measures only the charge deposited in one pixel.
\section{Conclusion}\label{sec:conclusion}
The threshold method has been successfully employed to precisely measure single pixel charge spectra from radioctive sources and in particle beams. It was demonstrated that after corrections for the systematic errors the charge resolution is limited by the electronic noise. In particular, the deficiency of the ATLAS pixel read out chip FE-I4 to provide good charge information has been overcome with a charge resolution better than \el{200}.
\acknowledgments
We want to thank Dr. Daniel Elsner, Dr. Frank Frommberger, Dr. Wolfgang Hillert and \mbox{Prof. F. Klein} for delivering high quality electron beams and their continous support at the ELSA test beams.
The research leading to these results has received funding from the European Commission under the FP7 Research Infrastructures project AIDA, grant agreement no. 262025 and from the German Federal Ministry of Education and Research (BMBF) under contract no. 05H2012PD1.\\
The information herein only reflects the views of its authors and not those of the European Commission and no warranty expressed or implied is made with regard to such information or its use.

\end{document}